\documentclass[12pt]{article}
\usepackage{graphicx,amsmath,hyperref}
\def\hybrid{\topmargin 0pt      \oddsidemargin 0pt
        \headheight 0pt \headsep 0pt
        \voffset=-0.5cm
        \textwidth 6.25in       
        \textheight 9.5in       
        \marginparwidth 0.0in
        \parskip 5pt plus 1pt   \jot = 1.5ex}
\catcode`\@=11
\def\marginnote#1{}

\newcount\hour
\newcount\minute
\newtoks\amorpm
\hour=\time\divide\hour by60
\minute=\time{\multiply\hour by60 \global\advance\minute by-\hour}
\edef\standardtime{{\ifnum\hour<12 \global\amorpm={am}%
        \else\global\amorpm={pm}\advance\hour by-12 \fi
        \ifnum\hour=0 \hour=12 \fi
        \number\hour:\ifnum\minute<10 0\fi\number\minute\the\amorpm}}
\edef\militarytime{\number\hour:\ifnum\minute<10 0\fi\number\minute}

\def\draftlabel#1{{\@bsphack\if@filesw {\let\thepage\relax
   \xdef\@gtempa{\write\@auxout{\string
      \newlabel{#1}{{\@currentlabel}{\thepage}}}}}\@gtempa
   \if@nobreak \ifvmode\nobreak\fi\fi\fi\@esphack}
        \gdef\@eqnlabel{#1}}
\def\@eqnlabel{}
\def\@vacuum{}
\def\draftmarginnote#1{\marginpar{\raggedright\scriptsize\tt#1}}
\def\draftlabel#1{{\@bsphack\if@filesw {\let\thepage\relax
   \xdef\@gtempa{\write\@auxout{\string
      \newlabel{#1}{{\@currentlabel}{\thepage}}}}}\@gtempa
   \if@nobreak \ifvmode\nobreak\fi\fi\fi\@esphack}
        \gdef\@eqnlabel{#1}}
\def\@eqnlabel{}
\def\@vacuum{}
\def\draftmarginnote#1{\marginpar{\raggedright\scriptsize\tt#1}}

\def\draft{\oddsidemargin -.5truein
        \def\@oddfoot{\sl preliminary draft \hfil
        \rm\thepage\hfil\sl\today\quad\militarytime}
        \let\@evenfoot\@oddfoot \overfullrule 3pt
        \let\label=\draftlabel
        \let\marginnote=\draftmarginnote
   \def\@eqnnum{(\theequation)\rlap{\kern\marginparsep\tt\@eqnlabel}%
\global\let\@eqnlabel\@vacuum}  }

\def\numberbysection{\@addtoreset{equation}{section}
        \def\theequation{\thesection.\arabic{equation}}}

\def\underline#1{\relax\ifmmode\@@underline#1\else
        $\@@underline{\hbox{#1}}$\relax\fi}

\def\titlepage{\@restonecolfalse\if@twocolumn\@restonecoltrue\onecolumn
     \else \newpage \fi \thispagestyle{empty}\c@page\z@
        \def\thefootnote{\fnsymbol{footnote}} }

\def\endtitlepage{\if@restonecol\twocolumn \else  \fi
        \def\thefootnote{\arabic{footnote}}
        \setcounter{footnote}{0}}  
\relax

\hybrid

\newtheorem{prop}{Proposition}[section]
\newtheorem{theor}[prop]{Theorem}

\newfont{\Bbb}{msbm10 scaled 1\@ptsize00}
\newfont{\Bbbb}{msbm7 scaled 1\@ptsize00}
\newcommand{\CC}{\mbox{\Bbb C}}

\newcommand{\DDD}{\raise-1pt\hbox{$\mbox{\Bbbb D}$}}

\newcommand{\UUU}{\raise-1pt\hbox{$\mbox{\Bbbb U}$}}

\newcommand{\ZZ}{\mbox{\Bbb Z}}
\newcommand{\z}{\raise-1pt\hbox{$\mbox{\Bbbb Z}$}}

\def\beq{\begin{equation}}
\def\eeq{\end{equation}}
\def\p{\partial}

\def\gtw{g_{0}}

\begin{document}

\begin{titlepage}

\title{Quantum Gaudin model and \\classical KP hierarchy}

\author{A. Zabrodin
\thanks{Institute of Biochemical Physics,
4 Kosygina, 119334, Moscow, Russia; ITEP, 25 B. Cheremushkinskaya,
117218, Moscow, Russia; National Research University Higher
School of Economics,
20 Myasnitskaya Ulitsa, Moscow 101000, Russia}}

\date{October 2013}
\maketitle

\vspace{-7cm} \centerline{ \hfill ITEP-TH-38/13}\vspace{7cm}

\begin{abstract}

This short note is a review of the intriguing 
connection between the quantum Gaudin model and the classical 
KP hierarchy recently established in \cite{ALTZ13}.
We construct the generating function of integrals 
of motion for the quantum Gaudin model with twisted 
boundary conditions (the master $T$-operator) and show that it satisfies
the bilinear identity and Hirota equations 
for the classical KP hierarchy. This implies that
zeros of eigenvalues of the master $T$-operator in the 
spectral parameter have the same dynamics as the Calogero-Moser
system of particles.

\end{abstract}

\end{titlepage}

\vspace{5mm}

\vspace{5mm}


{\small
\section{Introduction}

In \cite{ALTZ13}, a remarkable correspondence between 
the quantum Gaudin model and the classical Kadomtsev-Petviashvili (KP) hierarchy
was established. It is a limiting case of
the correspondence between quantum spin chains
with the Yangian $Y(gl(N))$ symmetry algebra 
and the classical modified
KP (mKP) hierarchy based on the construction of the master $T$-operator
\cite{AKLTZ11,Z12}. 
The master $T$-operator was
introduced in \cite{AKLTZ11} (in a preliminary form, it
previously appeared in \cite{KLT10}).
It is a special
generating function for commuting integrals of motion in the 
quantum model. In the Gaudin model case, any eigenvalue 
of the master $T$-operator 
appears to be the tau-function 
of the KP hierarchy, with polynomial dependence
on the spectral parameter. Taking into account 
the well known story about
dynamics of poles of rational solutions to soliton equations
\cite{AMM,Krichever-rat,Shiota}, 
this implies a link between the 
quantum Gaudin model \cite{Gaudin} and the classical 
Calogero-Moser (CM) system of particles \cite{Calogero}.
This link was also observed earlier in \cite{MTV}
using different arguments.

In this paper we present the results of \cite{ALTZ13} in a 
short compressed form. All proofs and technical details are omitted. 
Here we outline the results reviewed in the paper.
Most of them can be obtained by a limiting procedure
from the corresponding results for quantum spin chains proved in 
\cite{AKLTZ11}.

Using the matrix derivative operation,
we construct commuting integrals of motion for the $gl(N)$ Gaudin model,
with twisted 
boundary conditions and 
vector representations at the marked points in the quantum space,
corresponding to arbitrary representations in the auxiliary 
space. 
The master $T$-operator is their generating function.
It depends on an infinite number
of auxiliary ``time variables'' ${\bf t}=\{t_1, t_2, t_3 , \ldots \}$,
where $t_1$ can be identified with the spectral parameter $x$.
The master $T$-operator satisfies the bilinear identity
for the classical KP hierarchy and hence
any of its eigenvalues is a KP tau-function \cite{Sato,DJKM83}. 
This is a development of earlier studies \cite{KLWZ97,Z97,KSZ08}
clarifying the role of 
the Hirota bilinear equations \cite{Hirota81}
in quantum integrable models.

Moreover, all eigenvalues of the master $T$-operator are
{\it polynomial} tau-functions in $x=t_1$. Therefore, according to
\cite{Krichever-rat,Shiota}, the dynamics of their 
roots in $t_i$ with $i>1$ is
given by equations of motion of the classical CM system
of particles. 
The marked points $x_i$ in the Gaudin model (the inhomogeneities
at the sites in the spin chain language) should be identified with
initial coordinates of the CM particles while eigenvalues
of the Gaudin Hamiltonians are proportional to their initial momenta. 
Eigenvalues of the Lax matrix for the CM model coincide with
eigenvalues of the twist matrix (with certain multiplicities).
Therefore, with fixed integrals of motion
in the CM model determined by invariants of the
twist matrix, there are finite number of solutions for the values
of initial momenta
which correspond to different eigenstates of the Gaudin model.
In other words, the eigenstates of the Gaudin Hamiltonians 
are in one-to-one correspondence with (a finite number of) 
intersection points
of two Lagrangian submanifolds in the phase space of the 
classical CM model. This ``quantum-classical correspondence''
was also discussed in \cite{NRS,GK13} in the context of 
supersymmetric gauge theories and branes.

In distinction to paper \cite{ALTZ13}, here we consider 
the Gaudin model with a formal Planck's constant $\hbar$.
In this case the master $T$-operator is the tau-function
of the $\hbar$-dependent version of the KP hierarchy \cite{TakTak}
and the coupling constant of the CM model becomes proportional
to $\hbar^2$.

\section{The quantum Gaudin model}

Let ${\sf e}_{ab}^{\hbar}$ be generators of the 
``$\hbar$-dependent version'' of the universal 
enveloping algebra
$U(gl(N))$ with the commutation relations
$[{\sf e}^{\hbar}_{ab}, \,{\sf e}^{\hbar}_{a'b'}]=
{\hbar}(\delta_{a'b}{\sf e}^{\hbar}_{ab'}-
\delta_{ab'}{\sf e}^{\hbar}_{a'b})$.
The parameter $\hbar$ will play the role of the Planck's constant.
Let $\pi_{\lambda}$ be the finite dimensional irreducible
representation 
of $U(gl(N))$
with the highest weight $\lambda$.
We identify $\lambda$ with the Young diagram $\lambda =
(\lambda_1, \lambda_2 , \ldots , \lambda_{\ell})$ with
$\ell=\ell(\lambda )$ non-zero rows, where
$\lambda_i \in \ZZ_{+}$,
$\lambda_1 \geq \lambda_2 \geq \ldots \geq \lambda_{\ell}>0$.
For example, $\pi_{(1)}$ is the $N$-dimensional vector representation
corresponding to the 1-box diagram $\lambda =(1)$.
We have $\pi_{(1)}({\sf e}^{\hbar}_{ab})=\hbar e_{ab}$, where
$e_{ab}$ is the standard
basis in the space of $N\! \times \! N$ matrices:
the matrix $e_{ab}$ has only
one non-zero element (equal to 1) at the place $ab$:
$(e_{ab})_{a'b'}=\delta_{aa'}\delta_{bb'}$. 
Note that $I=\sum_a e_{aa}$ is the unity matrix and
$P=\sum_{ab}e_{ab} \otimes e_{ba}$ is the permutation operator
acting in the space $\CC^N \otimes \CC^N$.

In the tensor product $U(gl(N))^{\otimes n}$ 
the generators
${\sf e}^{\hbar}_{ab}$
can be realized as
${\sf e}^{\hbar \, (i)}_{ab}:= \mbox{id}^{\otimes (i-1)} \otimes
{\sf e}^{\hbar}_{ab} \otimes  \mbox{id}^{\otimes (n-i)}$.
Clearly, they commute for  any $i \ne j$ and any $a,b$ 
because act non-trivially in different spaces. 
Similarly, for any matrix $g\in \mbox{End}(\CC^N)$ we 
define $g^{(i)}$ acting in the tensor product 
${\cal V}=(\CC^N)^{\otimes n}$:
$g^{(i)}= I^{\otimes (i-1)} \otimes
g\otimes  I^{\otimes (n-i)}\in \mbox{End}({\cal V})$.
In this notation, 
$\displaystyle{P_{ij}:=\sum_{a,b}e^{(i)}_{ab}e^{(j)}_{ba}}$ is the
permutation operator of the $i$-th and $j$-th tensor factors
in ${\cal V}=\CC^N \otimes \ldots \otimes \CC^N$.

Fix $n$ distinct numbers $x_i\in \CC$ and
a diagonal $N\! \times \! N$
matrix $\gtw =\mbox{diag}\,(k_1, \ldots , k_N)$.
(We assume that the $k_i$'s are all distinct and non-zero.)
We will call $\gtw$ the twist matrix. 
The commuting Gaudin Hamiltonians are
\beq\label{gaudin-gen}
H_i=\sum_{a=1}^{N}k_a {\sf e}^{\hbar \, (i)}_{aa} +
\sum_{j\neq i}\sum_{a,b=1}^{N}
\frac{{\sf e}^{\hbar \,(i)}_{ab}{\sf e}^{\hbar \,(j)}_{ba}}{x_i-x_j}\,, 
\quad i=1,\ldots , n.
\eeq
The Hamiltonians of the quantum 
Gaudin model \cite{Gaudin} with the Hilbert space 
${\cal V}=(\CC^N)^{\otimes n}$ are restrictions of the operators 
(\ref{gaudin-gen}) to the $N$-dimensional vector representation:
\begin{equation}\label{D4}
H_i=\hbar \sum_{a=1}^{N}k_a e^{(i)}_{aa} +
\hbar^2 \sum_{j\neq i}\sum_{a,b=1}^{N}
\frac{e^{(i)}_{ab}e^{(j)}_{ba}}{x_i-x_j}=
\hbar \gtw ^{(i)} + \hbar^2 \sum_{j\neq i}\frac{P_{ij}}{x_i-x_j}
\,, \quad i=1,\ldots , n.
\end{equation}

It is easy to check that the operators
\beq\label{Ma}
M_a=\sum_{l=1}^{n}e_{aa}^{(l)}
\eeq
commute with the
Gaudin Hamiltonians: $[H_i, M_a]=0$.
Therefore, common eigenstates of the Hamiltonians can be classified
according to eigenvalues of the operators $M_a$. Let
$$\displaystyle{
{\cal V}=\bigotimes_{i=1}^{n}V_i \,\, =
\bigoplus_{m_1, \ldots , m_N} \!\! \!\! {\cal V}(\{m_a \})}
$$
be the decomposition of the Hilbert space of the Gaudin model
${\cal V}$ into the direct sum of eigenspaces for the operators
$M_a$ with the eigenvalues $m_a \in \ZZ_{\geq 0}$, $a=1, \ldots , N$.
Then the eigenstates of $H_i$'s are in the spaces ${\cal V}(\{m_a \})$.
Clearly,
$\sum_a M_a =n I ^{\otimes n}$, and hence
$\displaystyle{
\sum_{a=1}^{N}m_a =n}$.
Note also that
$
\displaystyle{\sum_{i=1}^{n}H_i= \sum_{a=1}^{N}k_a M_a}
$.

A more general family 
of commuting Gaudin Hamiltonians
was discussed in \cite{FFR,ER,Talalaev}. 
In \cite{ALTZ13}, it was shown that the $gl(N)$ Gaudin model with 
vector representations at the sites admits a 
very simple construction of the higher commuting Hamiltonians,
which is a version of the one suggested in
\cite{Kazakov2007na} for the spin chains of the $XXX$-type.
The main technical tool is the matrix derivative.

Let $g$ be an element of the Lie algebra $gl(N)$ and $f$ be any
function of $g$, with values in $\mbox{End}(L)$, where $L$ is the space
of any representation of $gl(N)$.
The matrix derivative is defined as follows:
\begin{equation}\label{d1}
{\sf d}f(g)=\frac{\p}{\p \varepsilon}\sum_{ab}e_{ab}\otimes
f(g+ \varepsilon {\sf e}_{ba})\Bigr |_{\varepsilon =0}.
\end{equation}
The right hand side belongs to $\mbox{End}\, (\CC^N \otimes L)$.
For example:
$$
{\sf d}\, (\mbox{tr} \, g)^k= k (\mbox{tr} \, g)^{k-1}I,
\quad
{\sf d}\, \mbox{tr} \, g^k =kg^{k-1}, \quad
{\sf d} g^k =
\displaystyle{P  \sum_{i=0}^{k-1}g^i\otimes g^{k-i-1}} 
\quad \mathrm{for} \quad k \in \ZZ_{\ge 0}
$$
When the number of tensor factors is more than two
another notation is more convenient.
Let $V_i\cong \CC^N$ be copies of $\CC^N$
and ${\cal V}=V_1\otimes \ldots \otimes V_n$ as before.
Then, 
applying the matrix derivatives to a scalar function $f$ 
several times, we can embed the result into $\mbox{End} ({\cal V})$
according to the formulas 
$$
{\sf d}_if(g)=\frac{\p}{\p \varepsilon}\sum_{ab}e_{ab}^{(i)}
f(g+ \varepsilon {\sf e}_{ba})\Bigr |_{\varepsilon =0},
$$ 
$$
{\sf d}_{i_2} {\sf d}_{i_1} f(g)=
\frac{\p}{\p \varepsilon_2}
\frac{\p}{\p \varepsilon_1}
\sum_{a_2b_2}\sum_{a_1b_1}
e^{(i_2)}_{a_2b_2}e^{(i_1)}_{a_1b_1}
f\left (
g+\varepsilon_1 {\sf e}_{b_1a_1}+
\varepsilon_2{\sf e}_{b_2a_2}\right )
\Bigr |_{\varepsilon_1=\varepsilon_2=0}
$$
and so on. The lower indices of ${\sf d}$ show in which tensor 
factors the resulting operator acts non-trivially.

Let $\chi_{\lambda}(g)=\mbox{tr}_{\pi_{\lambda}}g$ be
the character of the representation $\pi_{\lambda}$ calculated
for a matrix $g$. It 
is given in terms of the
Schur polynomials
$s_{\lambda}({\bf y})$
of the variables ${\bf y}=\{y_1, y_2, \ldots \}$,
$y_k=\frac{1}{k}\, \mbox{tr}\, g^k$:
\begin{align}
\chi_{\lambda}(g)=s_{\lambda}({\bf y})=
\det_{i,j=1, \ldots , \ell (\lambda )}h_{\lambda_i -i+j}({\bf y})
 \label{JT-det}
\end{align}
(the Jacobi-Trudi formula), with the complete symmetric polynomials
$h_{k}({\bf y})=s_{(k)}({\bf y})$ defined by
\begin{align}
\exp \Bigl (\xi({\bf y},z) \Bigr )=
\sum_{k=0}^{\infty}h_{k}({\bf y})z^{k},
\label{Schur-p}
\end{align}
where $\xi({\bf y},z) := \sum_{k\geq 1}y_kz^k$.
For example, $\chi_{(1)}(g)=\mbox{tr}\, g$.
It is convenient to set $h_{k}=0$ at $k<0$.
Let $p_1, \ldots , p_N$ be eigenvalues of $g$ realized
as an element of $\mbox{End} \, (\CC^N)$. Then
$y_k = \frac{1}{k}\, (p_1^k +\ldots +p_N^k)$ and
$\displaystyle{
\chi_{\lambda}(g)= \frac{\det_{1\leq i,j\leq N}
\bigl (p_j^{\lambda _i +N-i}\bigr )}{\det_{1\leq i,j\leq N}
\bigl (p_j^{N-i}\bigr )}}$
(see \cite{Macdonald}). This formula implies that
$\chi_{\emptyset}(g)=s_{\emptyset}({\bf y})=1$.
The characters form a special class of scalar functions on the space of 
$N\! \times \! N$ matrices which is of primary importance for us. 

Now we are ready to construct the
family of commuting operators for the Gaudin model:
\begin{equation}\label{d2}
{\sf T}^G_{\lambda}(x)=\Bigl ( 1+\frac{\hbar {\sf d}_n}{x-x_n}\Bigr )
\ldots \Bigl ( 1+\frac{\hbar {\sf d}_1}{x-x_1}\Bigr )
\chi_{\lambda}(\gtw )
\end{equation}
By the analogy with spin chains, we will call them Gaudin transfer-matrices.
The first few are: ${\sf T}^G_{\emptyset}(x)=1$,
$\displaystyle{{\sf T}^G_{(1)}(x)=
\mbox{tr}\, \gtw +\sum_i \frac{\hbar}{x-x_i}}$ and
\begin{equation}\label{g2}
\begin{array}{l}
\displaystyle{
{\sf T}^G_{(1^2)}(x)=\chi_{(1^2)}(\gtw )+\mbox{tr}\, \gtw
\sum_i \frac{\hbar}{x-x_i}+\sum_{i<j}\frac{\hbar^2}{(x-x_i)(x-x_j)}-
\sum_i \frac{H_i}{x-x_i}},
\end{array}
\end{equation}
From the last formula we see that the Gaudin Hamiltonians $H_i$ 
belong to this family. For one-column
diagrams, this construction agrees with Talalaev's prescription
\cite{Talalaev}.
At $n=0$ the transfer matrix is just the character:
$
{\sf T}^{G\, (n=0)}_{\lambda}(x)=\chi_{\lambda}(\gtw )
$.

In what follows, the normalization such that any Gaudin transfer 
matrix is a polynomial in $x$ will be more convenient.
In the polynomial normalization, the Gaudin 
transfer matrices are introduced by the formula
\begin{equation}\label{g1}
T^G_{\lambda}(x)=(x-x_n +\hbar {\sf d}_n)\, \ldots \,
(x-x_1 +\hbar {\sf d}_1)\chi_{\lambda}(\gtw ).
\end{equation}
All these operators commute for any $x$ and $\lambda$.

\section{The master $T$-operator and the KP hierarchy}

Let ${\bf t}=\{t_1, t_2, \ldots \}$ be an infinite set of ``time
parameters''. The master $T$-operator for the Gaudin model is the following 
generating function for the transfer-matrices $T^G_{\lambda}(x)$:
\begin{equation}\label{master2}
T^G(x, {\bf t})=\sum_{\lambda}T^G_{\lambda}(x)s_{\lambda}({\bf t}/\hbar ).
\end{equation}
These operators commute for different values of the parameters:
$[T^G(x,{\bf t}), \, T^G(x',{\bf t'})]=0.$
Since $\sum_{\lambda}\chi_{\lambda}(\gtw )s_{\lambda}({\bf t}/\hbar )=
\exp \Bigl (\frac{1}{\hbar}
\sum_{k\geq 1}t_k \, \mbox{tr}\, \gtw ^k\Bigr )$ (the 
Cauchy-Littlewoodn identity, see, e.g., 
\cite{Macdonald}), we can define the master $T$-operator
more explicitly as
\begin{equation}\label{master1}
T^G(x, {\bf t})=(x-x_n +\hbar {\sf d}_n)\, \ldots \,
(x-x_1 +\hbar {\sf d}_1)\exp 
\Bigl (\frac{1}{\hbar}\sum_{k\geq 1}t_k \, \mbox{tr}\, \gtw ^k\Bigr ).
\end{equation}
Note that because $e^{-t \, \mbox{\scriptsize{tr}}\, g}
{\sf d}e^{t \, \mbox{\scriptsize{tr}}\, g}=t$,
the role of the variable $t_1$
is to shift $x\to x+t_1$, so that 
$e^{\frac{1}{\hbar}x \, \mbox{\scriptsize{tr}}\, \gtw }T^G(x, {\bf t})$
depends on $x,t_1$ only through their sum $x+t_1$.

The master $T$-operator contains the complete information 
about the spectrum of all transfer-matrices. They can be restored
from it according to the formula
\begin{equation}\label{master3}
T^G_{\lambda}(x)=s_{\lambda}(\hbar \tilde \p )
T^G(x, {\bf t})\Bigr |_{{\bf t}=0},
\end{equation}
where $\tilde \p =\{ \p_{t_1}, \frac{1}{2}\, \p_{t_2},
\frac{1}{3}\, \p_{t_3}, \ldots \}$.
In particular,
\begin{equation}\label{master3a}
T^G_{(1)}(x)=\hbar \p_{t_1}T^G(x, {\bf t})\Bigr |_{{\bf t}=0}, \quad
T^G_{(1^2)}(x)=\frac{1}{2}\,
(\hbar ^2 \p_{t_1}^2 -\hbar \p_{t_2})T^G(x, {\bf t})\Bigr |_{{\bf t}=0}.
\end{equation}

For any $z\in \CC$ we put
$
{\bf t}\pm \hbar [z^{-1}]:= 
\Bigl \{t_1 \pm \hbar z^{-1}, \, t_2 \pm \frac{\hbar}{2}\,
z^{-2}, \, t_3 \pm \frac{\hbar}{3}\,
z^{-3}, \, \ldots \Bigr \}
$.
As we shall see below,
$T^G(x, {\bf t}\pm \hbar [z^{-1}])$ regarded as functions of
$z$ with fixed ${\bf t}$ play an important role.
Here we only note that equation (\ref{master3}) implies that
$T^G(x, 0\pm \hbar [z^{-1}])$ is the generating series for
$T$-operators corresponding to the one-row and one-column diagrams:
\begin{equation}\label{master4}
T^G(x,  \hbar [z^{-1}])=\sum_{s\geq 0}z^{-s}T^G_{(s)}(x), \quad \quad
T^G(x,  -\hbar [z^{-1}])=\sum_{a=0}^{N}(-z)^{-a}T^G_{(1^a)}(x).
\end{equation}

The following theorem is the main result of \cite{ALTZ13}.
\begin{theor}
The master $T$-operator (\ref{master1})
satisfies the bilinear identity for the 
$\hbar$-\-de\-pen\-dent KP hierarchy
\cite{Sato,DJKM83,TakTak}:
\begin{equation}\label{hir1}
\oint_{\infty} e^{\frac{1}{\hbar}\xi ({\bf t}-{\bf t'}, z)}\,
T^G\left (x, {\bf t}-\hbar [z^{-1}]\right )
T^G\left (x, {\bf t'}+\hbar [z^{-1}]\right )dz =0 \quad
\mbox{for all ${\bf t}, {\bf t}'$.}
\end{equation}
The integration contour is chosen in such a way that 
it encircles all singularities
coming from the $T^G$'s and none of those coming from 
$e^{\frac{1}{\hbar}\xi ({\bf t}-{\bf t'}, z)}$. 
\end{theor}

\noindent
This means that each eigenvalue of the master $T$-operator 
is a tau-function of the KP hierarchy. The general bilinear identity 
(\ref{hir1}) implies many bilinear funtional relations for 
the master $T$-operator (the Hirota equations). Some of them
are written explicitly in \cite{ALTZ13}.
Equation (\ref{master2}) is the expansion of the tau-function
in Schur polynomials \cite{Sato,Orlov-Shiota,EH}.

As soon as the Gaudin model is linked to the KP hierarchy,
it is tempting to ask about the role of the other standard 
ingredients of the KP theory. We will be mostly interested 
in the Baker-Akhiezer (BA) function.
According to the general scheme, the BA
function corresponding to the tau-function (\ref{master1})
is given by the formula
\beq\label{BA1}
\psi (x, {\bf t};z)=e^{\frac{1}{\hbar}(xz +\xi ({\bf t}, z))}
(T^G(x, {\bf t}))^{-1}\, T^G\bigl (x, {\bf t}-\hbar [z^{-1}]\bigr ).
\eeq
and satisfies the differential
equation
\beq\label{BA7}
\hbar \p_{t_2}\psi (x, {\bf t};z)=\hbar^2 \p_x^2 \psi (x, {\bf t};z) +
2u(x, {\bf t}) \psi (x, {\bf t};z),
\eeq
where
$u(x, {\bf t})=
\hbar^2 \p_x^2 \log T^G(x, {\bf t})$. 
Using the definition (\ref{master1}), we have:
\beq\label{BA3}
\psi (x, {\bf t};z)=z^{-N}e^{\frac{1}{\hbar}(xz +\xi ({\bf t}, z))}
(T^G(x, {\bf t}))^{-1} \!
\overleftarrow{\prod_{i=1}^{n}}
(x\! -\! x_i \! +\! \hbar {\sf d}_i)
\left [ \det \left (z I\! -\! \gtw \right )
\exp \Bigl (\frac{1}{\hbar}
\sum_{k\geq 1}t_k \, \mbox{tr}\, h^k\Bigr )\right ]
\eeq
(here and below 
$\displaystyle{\overleftarrow{\prod_{i=1}^{n}}A_i =A_{n}\ldots A_2A_1}$).
From this formula we see that
$e^{-\frac{1}{\hbar}(xz +\xi ({\bf t}, z))}
\psi (x, {\bf t};z)$ is a polynomial
in $z^{-1}$ of degree $N$. 
Regarded as a function of $x$, it 
is a rational function with
$n$ zeros and $n$ poles which are simple in general position.
From (\ref{master1}), (\ref{BA3}) it follows that
\beq\label{BA5}
\lim_{x\to \infty}
\Bigl (e^{-\frac{1}{\hbar}(xz +\xi ({\bf t}, z))}\psi (x, {\bf t};z)
\Bigr )=z^{-N}\det (zI-\gtw ).
\eeq
We also note the formula for the {\it stationary} BA
function $\psi (x,z):=\psi (x,0;z)$,
which directly follows from (\ref{BA3}):
\beq\label{BA3st}
\psi (x, z)=z^{-N}e^{\frac{1}{\hbar}xz}\,
\overleftarrow{\prod_{i=1}^{n}}
\Bigl (1+\frac{\hbar {\sf d}_i}{x-x_i}\Bigr )
\det (zI-\gtw ).
\eeq

\section{The link to the classical Calogero-Moser model}

Integrability of the Gaudin model implies that
eigenvalues of the master $T$-operator in the polynomial normalization are 
polynomials in the spectral parameter $x$ of degree $n$:
\begin{equation}\label{CM1}
T^G(x, {\bf t})=e^{\frac{1}{\hbar}
t_1 \, {\scriptsize{\mbox{tr}}\, \gtw }
+\frac{1}{\hbar} t_2 \, {\scriptsize{\mbox{tr}}\, \gtw ^2 +\ldots}}
\prod_{k=1}^{n}(x+t_1 -x_k(t_2, t_3, \ldots )).
\end{equation}
The roots of each eigenvalue have their own
dynamics in the times $t_k$. This dynamics is known \cite{Krichever-rat,Shiota}
to be given by the rational CM model \cite{Calogero}.
The parameters $x_i$ of the Gaudin model play the role of
coordinates of the CM particles at $t_i=0$:  $x_j=x_j(0)$.
In particular, we have
$T^G(x,0)=T^G_{\emptyset}(x)=\prod_{k=1}^{n}(x-x_k)$.

Using (\ref{master3a}), we easily obtain the formula
for ${\sf T}^G_{(1)}(x)= T^G_{(1)}(x)/T^G_{\emptyset}(x)$
given in (\ref{g2}). For ${\sf T}^G_{(1^2)}(x)$ the second
equation in (\ref{master3a}) yields
$$
{\sf T}^G_{(1^2)}(x)=\chi_{(1^2)}(\gtw )+
\mbox{tr}\, \gtw
\sum_i \frac{\hbar}{x-x_i}+\sum_{i<j}\frac{\hbar^2}{(x-x_i)(x-x_j)}
+\frac{1}{2}\sum_i \frac{\hbar \dot x_i}{x-x_i},
$$
where $\dot x_i =\p_{t_2}x_i(t_2)\Bigr |_{t_2=0}$.
Comparing with the third equation in (\ref{g2}), we conclude that
the initial velocities are proportional to eigenvalues of the
Gaudin Hamiltonians:
\begin{equation}\label{CM2}
\hbar \dot x_i =-2 H_i.
\end{equation}
This unexpected connection between the quantum Gaudin model 
and the classical CM model was observed in 
\cite{MTV} using different arguments.

Following 
\cite{Krichever-rat}, one can derive equations of motion
for the $t_k$-dynamics of the $x_i$'s performing the pole expansion
of the linear problem (\ref{BA7}) for the BA function $\psi$. 
It is convenient to denote $t_2=t$ and put all othet times equal 
to $0$ because they are irrelevant in this derivation.
The pole ansatz for the BA function is
\beq\label{CM4}
\psi = e^{\frac{1}{\hbar}(xz+t z^2)}\left (c_0 (z) +
\sum_{i=1}^{n}\frac{c_i(z,t)}{x-x_i (t)}\right ),
\eeq
where $c_0(z)=\det (I -z^{-1}\gtw )$ .
One should substitute it into the linear problem (\ref{BA7}) with
$\displaystyle{
u=-\sum_{i=1}^{n}\frac{\hbar^2}{(x-x_i)^2}}$
and cancel all the poles at the points $x_i$. This yields an
overdetermined system of equations for the coefficients $c_i$.
Their compatibility implies the
the Lax representation for the CM model:
\beq\label{CM8}
\dot Y=[T, \, Y],
\eeq
where the $n \! \times \!n$ matrices $Y$, $T$ are
given by
\beq\label{CM6}
Y_{ik}= -p_i \delta_{ik}-
\hbar \, \frac{1-\delta_{ik}}{x_i-x_k}\,, \quad \quad
p_i:=\frac{1}{2}\, \dot x_i,
\eeq
\beq\label{CM7}
T_{ik}=-\delta_{ik}\sum_{j\neq i}\frac{2\hbar}{(x_i-x_j)^2}+
\frac{2\hbar (1-\delta_{ik})}{(x_i-x_k)^2}.
\eeq
The equations of motion are:
\beq\label{CM9}
\ddot x_i = -8\sum_{j\neq i} \frac{\hbar^2}{(x_i-x_j)^3}.
\eeq
Set $X=X({\bf t})=\mbox{diag} (x_1({\bf t}), \ldots , x_n({\bf t}))$.
For the function $\psi$ itself we then have:
\beq\label{cc1}
\psi =\det (I-z^{-1}\gtw )\, e^{
\frac{1}{\hbar}(xz+tz^2+t_3 z^3 +\ldots )}\Bigl (
1-\hbar {\sf 1}^{\sf t}(xI -X)^{-1}(zI -Y)^{-1}{\sf 1}\Bigr ),
\eeq
where ${\sf 1} =(1, 1, \ldots , 1)^{\sf t}$ is the $n$-component
vector.
As is well known (and easy to check),
the matrices $X$, $Y$ satisfy the commutation
relation
\beq\label{comm}
[X, \, Y]=\hbar (I -{\sf 1}\otimes {\sf 1}^{\sf t})
\eeq
(here ${\sf 1}\otimes {\sf 1}^{\sf t}$ is the $n\! \times \! n$
matrix of rank $1$ with all entries equal to $1$).

The matrix $Y$ is the Lax matrix for the CM model. 
As is seen from (\ref{CM8}), the time evolution preserves
its spectrum, i.e., the coefficients 
${\cal J}_k$ of the characteristic polynomial
\beq\label{char-pol}
\det (zI -Y(t))=\sum_{k=0}^{n}{\cal J}_k z^{n-k}
\eeq
are integrals of motion. The highest integral, ${\cal J}_n$, 
was found explicitly in \cite{Sawada-Kotera}, where a recurrence 
procedure for finding all other integrals of motion was also 
suggested. 
In fact this procedure is equivalent to the following
explicit expression for the characteristic polynomial:
\beq\label{explicit}
\det \bigl (zI -Y(t)\bigr )=\exp \Bigl ( \sum_{i<j}
\frac{\hbar^2 \p_{p_i}\p_{p_j}}{(x_i-x_j)^2}\Bigr ) 
\prod_{l=1}^{n}(z+p_l).
\eeq
Note that this expression
is well-defined because the sum obtained after expansion 
of the exponential function in the r.h.s. contains a finite 
number of non-zero terms.

One can see that eigenvalues of the Lax matrix $Y$
are the same as eigenvalues of the twist matrix $\gtw$ (with appropriate
multiplicities). Indeed, let us compare expansions of (\ref{BA3st}) and
(\ref{cc1}) at large $x$. From (\ref{BA3st}) we have:
$$
\psi (x,z)=\det (I-z^{-1}\gtw ) \, e^{\frac{1}{\hbar}xz} \left (
1-\frac{\hbar}{x}\sum_i \sum_a \frac{e_{aa}^{(i)}}{z-k_a} +
O(x^{-2})\right ).
$$
Using the commutation relation (\ref{comm}) 
it is easy to check that for any $k\geq 0$ it holds 
$
{\sf 1}^{\sf t} Y^k {\sf 1} = \mbox{{\rm tr}}\, Y^k$.
Taking this into account, 
we can expand (\ref{cc1}) at ${\bf t}=0$:
$$
\psi (x,z)=\det (I-z^{-1}\gtw ) \, e^{\frac{1}{\hbar}xz} \left (
1-\frac{\hbar}{x}\, \mbox{tr} \,\frac{1}{z-Y_0} \, +O(x^{-2})\right ),
$$
where $Y_0:=Y(0)$.
Therefore, we conclude that
$\displaystyle{
\mbox{tr} \,\frac{1}{zI-Y_0} =
\sum_i \sum_a \frac{e_{aa}^{(i)}}{z-k_a}}
$
and, since $\mbox{tr}\, (zI-Y_0)^{-1}=\p_z \log \det (zI-Y_0)$, we have
\beq\label{cc2}
\det (zI-Y_0)=\prod_{a=1}^{N}(z-k_a)^{\sum_{i=1}^{n}e_{aa}^{(i)}}
=\prod_{a=1}^{N}(z-k_a)^{M_a},
\eeq
where $M_a$ is the operator (\ref{Ma}). Hence we see that
the $M_a$ is the ``operator multiplicity'' of the
eigenvalue $k_a$. In the sector ${\cal V}(\{m_a\})$ the
multiplicity becomes equal to $m_a$.
This argument allows one to prove the following important statement:
\begin{theor}
The eigenvalues of the Lax matrix $Y$ of the CM model are numbers from the set
$\{k_1, k_2, \ldots , k_N\}$ (the eigenvalues of the twist matrix) with multiplicities $m_a\geq 0$ such that
$m_1 +\ldots +m_N =n$, with the $m_a$'s being eigenvalues of the 
operators $M_a$.
\end{theor}

The hamiltonian form of equations of motion (\ref{CM9}) is
$\displaystyle{
\left (\begin{array}{l}\dot x_i \\ \dot p_i \end{array} \right )=
\left (\begin{array}{r} \p_{p_i} {\cal H}_2 \\
- \p_{x_i} {\cal H}_2 \end{array} \right )}$
with the Hamiltonian
\beq\label{CM11}
{\cal H}_2= \mbox{tr}\, Y^2 = \sum_i p_i^2 -
\sum_{i<j}\frac{2\hbar^2}{(x_i-x_j)^2}\,.
\eeq
This result can be extended to the whole hierarchy \cite{Shiota}:
\beq\label{CM12}
\left (\begin{array}{l}\p_{t_k}x_i \\ \p_{t_k} p_i \end{array} \right )=
\left (\begin{array}{r} \p_{p_i} {\cal H}_k \\
- \p_{x_i} {\cal H}_k \end{array} \right ),
\quad \quad {\cal H}_k =\mbox{tr}\, Y^k.
\eeq
The ${\cal H}_k$'s are higher integrals of motion 
for the CM
model. They are known to be in involution 
\cite{Sawada-Kotera,Woj,OP}. This agrees with commutativity 
of the KP flows. The integrals ${\cal H}_k$ are connected with the
integrals ${\cal J}_k$ introduces in (\ref{char-pol}) by Newton's
formula \cite{Macdonald}
$\displaystyle{
\sum_{k=0}^{n}{\cal J}_{n-k}{\cal H}_k =0}
$ (we have set ${\cal H}_0 \! =\mbox{tr} Y^0 \! =\! n$).

The results of \cite{Shiota} imply an explicit
determinant representation of the tau-function.
It is easy to adopt it for the master $T$-operator $T^G(x, {\bf t})$
(\ref{CM1}). Let $X_0 =X(0)$ be the diagonal matrix
$X_0=\mbox{diag}(x_1, x_2, \ldots , x_n)$,
where $x_i=x_i(0)$ and $Y_0$ be the
Lax matrix (\ref{CM6}) at ${\bf t}=0$, with the diagonal elements
being 
proportional to the Gaudin Hamiltonians $H_i =-\hbar p_i (0)$:
\beq\label{S1}
Y_0= \left ( \begin{array}{ccccc}
\displaystyle{\frac{H_1}{\hbar}} & \displaystyle{\frac{\hbar}{x_2-x_1}} 
&\displaystyle{\frac{\hbar}{x_3-x_1}} &
\ldots & \displaystyle{\frac{\hbar}{x_n-x_1}}
\\ &&&& \\
 \displaystyle{\frac{\hbar}{x_1-x_2}} & 
 \displaystyle{\frac{H_2}{\hbar} }& 
 \displaystyle{\frac{\hbar}{x_3-x_2}} &
 \ldots & \displaystyle{\frac{\hbar}{x_n-x_2}}
 \\ &&&& \\ \vdots & \vdots & \vdots & \ddots & \vdots
 \\ &&&& \\
 \displaystyle{\frac{\hbar}{x_1-x_n}} & 
 \displaystyle{\frac{\hbar}{x_2-x_n}}&
 \displaystyle{\frac{\hbar}{x_3-x_n}} & \ldots & 
 \displaystyle{\frac{H_n}{\hbar}}
\end{array}\right ).
\eeq
\begin{theor}
The master $T$-operator for the Gaudin model is given by
\beq\label{CM13}
T^G(x, {\bf t})=e^{\frac{1}{\hbar}
\sum_{k\geq 1}t_k{\scriptsize{\mbox{{\rm tr}}}\, h^k}}
\det \left ( xI -X_0 +\sum_{k\geq 1}kt_k Y_0^{k-1}\right ).
\eeq
\end{theor}

It follows from the above arguments that eigenvalues of the
Gaudin Hamiltonians $H_i$, $i=1, \ldots , n$ (\ref{D4}),
can be found in the framework of the classical
CM system with $n$ particles. Namely,
the spectrum of $H_i$'s in the space ${\cal V}(\{ m_a\})$ is determined
by the conditions
\beq\label{S2}
\mbox{tr}\, Y_0^j = \sum_{a=1}^{N} m_a k_a^j \quad
\mbox{for all} \quad j\geq 1,
\eeq
i.e., given the initial coordinates $x_i$ and the action variables
${\cal H}_j =\mbox{tr}\, Y_0^j$ one has to find possible values
of the initial
momenta $p_i =-H_i/\hbar$. 
Taking into account equations (\ref{char-pol}) and (\ref{explicit}),
we can represent the equations for $H_i$ in the form of the equality
\beq\label{S3}
\exp \Bigl (\hbar^4 \sum\limits_{i<j}x_{ij}^{-2}
\p_{H_i}\p_{H_j}\Bigl )\prod_{l=1}^{n}(z-\hbar^{-1}H_l)=
\prod_{a=1}^{N}(z-k_a)^{m_a}\,, \quad \quad x_{ij}\equiv x_i-x_j
\eeq
which has to be satisfied identically in $z$. This identity
is equivalent to $n$ algebraic equations for $n$ quantities
$H_1, \ldots , H_n$.

We see that
the eigenstates of the Gaudin Hamiltonians correspond to 
the intersection points of two Lagrangian submanifolds:
one obtained by fixing the $x_i$'s and the other obtained by 
fixing the ${\cal H}_i$'s, with values of the latter being 
determined by eigenvalues of the twist matrix.
This purely classical prescription appears to be
equivalent to the Bethe ansatz solution and 
solves the spectral problem for the quantum Gaudin Hamiltonians.

{\bf Example.} Consider the vector ${\sf v}_a \in \CC^N$ with
components $({\sf v}_a)_b =\delta_{ab}$. Since
$P_{ij} ({\sf v}_a)^{\otimes n} = ({\sf v}_a)^{\otimes n}$, the vector
$({\sf v}_a)^{\otimes n}$ is an eigenstate for the Gaudin
Hamiltonians $H_i$ with the eigenvalues
$\displaystyle{k_a +\sum_{j\neq i}\frac{\hbar}{x_i-x_j}}$.
It is also an eigenvector for the operators $M_b$ with
eigenvalues $m_b=n\delta_{ab}$. The matrix (\ref{S1}) in this case
is the $n\! \times \! n$
Jordan block with the only eigenvector ${\sf 1}$ with eigenvalue
$k_a$ and $\mbox{tr}\, Y_0^j =nk_a^j$.

\section*{Acknowledgments}

Discussions with A.Alexandrov,
A.Gorsky, V.Kazakov, S.Khoroshkin, I.Krichever, S.Leu\-rent,
M.Ol\-sha\-nets\-ky, A.Orlov, T.Ta\-ke\-be, Z.Tsuboi, and A.Zo\-tov 
are gratefully acknowledged. 
Some of these results were reported 
at the International conference ``Physics and Mathematics of Nonlinear Phenomena'' (22-29 June 2013, Gallipoli, Italy).
The author thanks the organizers and especially professor
B.Konopelchenko for the invitation and support.
This work was supported in part
by RFBR grant 12-01-00525, by joint RFBR grants 12-02-91052-CNRS,
12-02-92108-JSPS and
by Ministry of Science and Education of Russian Federation
under contract 8207 and by
grant NSh-3349.2012.2 for support of
leading scientific schools.

}

\end{document}